\documentstyle[aps,prl,multicol,epsf]{revtex}
\draft
\begin{document}
 
 
\title{Josephson current in unconventional
 superconductors through an
Anderson impurity
\\
}
 
\author{ Y. Avishai$^{1,2}$ \cite{email1} and A. Golub$^{1}$ \cite{email} }
\address{$^1$Department of Physics, Ben-Gurion University of the Negev,
Beer-Sheva, Israel\\
$^2$Institute for Solid State Physics, University of Tokyo, 7-22-1 
Roppongi, Minato-ku, Tokyo 106, Japan}

\date{\today}
 
\maketitle
 
\begin{abstract}
Josephson current for a system consisting of
an Anderson impurity weakly coupled 
to two unconventional superconductors is studied and
shown to be driven by a
 surface zero energy (mid-gap) bound-state.
The repulsive Coulomb interaction in the dot can turn
a $\pi$ junction into a
 $0$ -junction. This effect is more pronounced in
$p$-wave superconductors while
in high-temperature superconductors  
 with $d_{x^2-y^2}$ symmetry it can exit for  rather large 
artificial centers at which tunneling occurs within a finite region.  
\end{abstract}

\begin{multicols}{2}
\narrowtext

Recently, the Josephson effect in unconventional superconductors 
have attracted a considerable attention \cite{hu,tan1,bar,tan2,sam,rie}. 
Measurements of direct Josephson current yield valuable
information on the symmetry of the order parameter which is essential for
understanding the mechanisms of superconductivity in these complex
materials. Phase interference experiments \cite{van} 
 definitely suggest the presence 
of $ d $-wave symmetry of the order parameter in high-$T_c$ superconductors,
while the recent discovery of 
superconductivity in  $ Sr_{2}RuO_{4} $ \cite {maeno} 
implies the existence of a peculiar system for which the pair
potential has a triplet ($p$-wave) symmetry \cite {rice}.

$p$ and $d$-wave symmetries of the order parameter have in
common a property which reflects the variation of the pair 
potential across the Fermi surface. This results in 
a strong sensitivity to
inhomogeneities which, in turn, influences the Josephson
effect.
In particular, an anomalous
 temperature dependence of a single Josephson
junction at low temperatures
\cite{bar,tan2,sam,rie} and an induced crossover from a usual
($0$-junctions ) to a $\pi$-junction on approaching the critical
temperature were observed.

Tunneling in a Josephson junction consisting of conventional
($s$-wave) superconductors and a dynamical
impurity (Anderson and Kondo ) was considered sometime ago
 \cite{soda,glas,spi,gol}. In a recent work \cite{aro} the tunneling 
current was calculated at zero temperature and was shown to be 
strongly dependent on the Coulomb
interaction which, in some cases, may cause a sign change of the current.
The experimental observations described above necessarily suggest 
that one must now go further and 
study the same device albeit with unconventional 
superconductors at finite temperatures. This missing part is addressed below.
As will be demonstrated, the underlying physics is remarkably different. 

The main focus here is the influence of Coulomb 
interaction on the low temperature behavior of 
the current in a 2D Josephson junction
consisting of left (L) and right (R) superconductors 
with either $p$ or $d_{x^2-y^2}$
symmetry of the order parameter) weakly coupled 
to a quantum dot ({\em via} 
identical hoping matrix elements $t_L=t_R=t$). 
The dot is represented by
a finite $U$ Anderson impurity
whose energy $-\epsilon_{0}<0$ lies below the Fermi energy.
Usually, the inequalities
$U>-\epsilon_{0}>0$ are maintained 
so that the ground state  of the disconnected ($t=0$)
dot is singly occupied.
We use the non-perturbative scheme suggested in Ref.\cite{aro} 
(extended for finite temperature) and elucidate the low temperature 
behavior of the Josephson current and its dependence on 
$U$, $t$ and the phase of the order parameter $\Delta$.

Since $\Delta$ is not isotropic, it is useful 
at this point to specify the underlying geometry. 
Each superconductor has the shape of half a plane defined as 
$-\infty < y < \infty$ and $x<0$ ($x>0$) for the left (right) 
superconductor. The dot is located at the origin ${\bf r}=0$ and 
tunneling is described by zero-range hoping between the impurity and 
the superconductors along the $x$ axis. 

For $d$-wave
superconductors we choose the nodal line of the pair 
potential on the Fermi surface
to coincide with the tunneling direction, such that
$\Delta=v_{\Delta}p_{F}\sin2\alpha$ where $p_F$ is 
the Fermi momentum. 
For spin-triplet
superconducting states the order parameter is an odd vector
function of momentum and a $ 2\times 2$ matrix in spin space.
We chose  to represent it by 
the time reversal symmetry breaking 
state\cite{rice} which is off-diagonal in spin
indices, that is, 
$\Delta=\Delta_{0}\exp i\alpha$. This state
 is a probable candidate for describing the 
recently discovered 
superconductor $Sr_{2}RuO_{4}$ \cite {maeno}.

In both cases, the pair potential
has different values
for electron like excitations which move at an angle $\alpha$ and hole
like excitations propagating along the direction $\pi-\alpha$. This fact
significantly affects the scattering process and causes the formation of a 
zero energy (mid-gap) bound state centered
at the boundary.

To obtain  the Josephson current we compute the partition
function $Z=\int D[\bar{\psi}\psi\bar{c}c]\exp(-S)$ 
and the corresponding free energy of the system. 
The functional integration
 is performed 
over Grassmann fields in the superconductors 
and the quantum dot (see below for a precise definition).
The Euclidean action can be written as a sum 
$S=S_{L}+S_{R}+S_{int}+S_{U}$, 
where, with obvious notations, 
\begin{equation}
S_{L}=T\sum_{\omega}\int dxdy\bar{\psi}_{L\omega}(xy)
(i\omega+
H^{BDG}_{L}(\hat{p}_{x},\hat{p}_{y})\psi_{L\omega}(xy),
\end{equation}
\begin{equation}
S_{int}=-T\sum_{\omega i=L,R}(t_{i}\bar{\psi}_{\omega i}(0)\tau_{3}
c_{\omega}+t^*_{i}\bar{c_{\omega}}\tau_{3}\psi_{\omega i}(0)),
\end{equation}
\begin{equation}
S_{U}=\int
d\tau[\bar{c}\partial_{\tau}c+\tilde{\epsilon}\bar{c}\tau_{3}c-
U(\bar{c}c)^{2}/2],
\end{equation}
where $ \tilde{\epsilon}=\epsilon_{0}+U/2$.
The summation is taken over odd Matsubara frequencies
$\omega=(2n+1)\pi T$, while $\bar{\psi}_{\omega i}\equiv
 (\bar{\psi}_{\omega i \uparrow} \psi_{\omega i \downarrow})
, \bar{c}\equiv(\bar{c}_{\uparrow}c_{\downarrow})$ 
and the corresponding
conjugate fields are Grassmann
variables of the superconductors and the impurity respectively.
  
The Bogolubov DeGennes hamiltonian $H^{BDG}_{i=L,R}$
acquires the form:
\[H^{BDG}_{i}=
 \left( \begin{array}{cc}
\epsilon(\hat{p})-p^2_{F}/2m & \Delta(\hat{p})\exp(\phi_{i})  \\

\Delta(\hat{p})^{*}\exp(-\phi_{i}) & -\epsilon(\hat{p})+p^2_{F}/2m 
\end{array}\right).\]
where $\phi_i$ are the phases of superconducting condensates,
and $\epsilon(\hat{p}) $ denotes the kinetic energy operator with dispersion
$\epsilon(p) $. Note that we take $\Delta_L=\Delta_R$ but $\phi_L \ne
\phi_R$

Since the pair potential is
translation invariant in the $y$ direction
we can express the action in terms of
fermion Grassmann variables $\psi_{\omega k_{y}}(x)$ and BDG Hamiltonian 
$H^{BDG}(\hat{p}_{x},k_{y})$ where $k_{y}=p_{F} sin \alpha$.
Moreover, due to the zero-range nature of the hopping, 
the fields at ${\bf r} \ne 0$ can be integrated out 
yielding effective actions in terms of the boundary 
fields $\psi_{\omega k_{y}}(0)$ alone\cite{aro}. Explicitly,
\begin{equation}
S_{L}=T\sum_{ \omega,k_{y}}\bar{\psi}_{L \omega k_{y}}(0)
G_{Lk_{y}}\psi_{L \omega k_{y}}(0),
\end{equation}
with,
\begin{equation}
G_{Lk_{y}}=\frac{k_{0}^{4}}{4m^2}\int_{-\infty}^{0}
\int_{ -\infty}^{0}dxdx'\tau_{3}
G_{Lk_{y}}(xx'),
\end{equation}
where $k_{0}=\sqrt{p^{2}_{F}-k_{y}^{2}}$. The
Green function $G_{Lk_{y}}(xx')$
satisfies the equation
$-\tau_{3}[i\omega+H^{BDG}_{L}(\hat{p}_{x},k_{y})]
G_{Lk_{y}}(xx')=\delta(x-x')$ with
homogeneous boundary conditions.
It is constructed as a sum over the basic solutions
of the equation 
$-\tau_{3}[i\omega+H^{BDG}_{L}
(\hat{p}_{x},k_{y})]u_{Lk_{y}}(x)=0$ with 
coefficients depending on $x'$. The spinors $u_{Lk_{y}}(x)$
describe electron like and hole like excitations moving
in different pair potentials. 
For $p$-wave 
superconductors one has $G_{Lk_{y}}=
\frac{k_{0}}{2m\omega}r_{L}(\omega)$ with 
\begin{equation}
r_{L}(\omega)=\left (\begin{array}{cc}
-i\sqrt{\omega^2 +|\Delta_{L}|^2} & \Delta_{L} \\
-\Delta^{*}_{L} & -i\sqrt{\omega^2 +|\Delta_{L}|^2}
\end{array} \right ). 
\end{equation}
It is markedly different from the corresponding quantity derived 
for $s$-wave superconductors (note in particular the occurrence of 
$\omega$ in the denominator).
Performing the integration over the boundary
fields
$\bar{\psi}_{L\omega k_{y}}(0),
\psi_{L\omega k_{y}}(0)$ in the
partition
function is now straightforward. The
remaining integrations over the $c,\bar c$ fields 
is done by decoupling the
Hubbard interaction using a Hubbard-Stratonovich transformation. 
Since the impurity operators do not depend on $k_{y}$, the 
auxiliary fields $\gamma$ depend
only on $\omega$. We then obtain

\begin{equation}
Z=C\int\prod_{\omega}d\gamma_{\omega}\exp(-\frac{\gamma_{\omega}^2}{2UT}-
\frac{\tilde{\epsilon}}{T}-\frac{Q}{T}). 
\end {equation}
Here $Q=-T\sum_{\omega} ln [detR(\omega)]$ and $ C $ is a constant. 
The 
matrix R encodes the coupling between the superconducting (half) planes
connected and the impurity, and reads
\begin{equation}
R(\omega)=\tilde{\epsilon}\tau_{3}+\gamma_{\omega}-
i\omega-\frac{\pi N(0)}{\omega}<|t|^2 (r_{L}(\omega)+
r_{R}(\omega))>,
\end {equation}
where $N(0)$ is the density of states at the Fermi level, and
$<O>\equiv\pi^{-1} \int_{-\pi/2}^{\pi/2}d\alpha O(\alpha)$. 
The averaging procedure thus defined
is related to a possible dispersion
of the transmission matrix element $t_{k{y}}$. For a point junction
with constant $t$, only the component of the Josephson current
perpendicular to the interface is relevant.
In this case, for even ($d$-wave symmetry) superconductors 
the mid-gap
zero energy bound state does not
contribute to the Josephson energy.
Contrary, for $ p$-wave pair potential this state defines
the main contribution at low temperatures \cite{tan4} (which is the 
temperature domain of our interest here).
Intuitively, occurrence of dispersive
tunneling matrix elements $t_{k{y}}$  correspond to deviation 
of the impurity from a point-like defect. Such a case
can be realized {\em e.g.} by
artificially - induced defects\cite{y}.
The spectroscopy of $ Bi_{2}Sr_{2}CaCu_{2}O_{8}$ surfaces
indicates that such defects appear to be more extended in STM imaging.
In this case one can expect non-zero contribution from the mid-gap level
in $d$-waves superconductors as well.

With this point in mind we now proceed and
consider the $p$-wave case. 
The functional integral in (7) is 
approximated by the saddle point method. 
The appropriate optimum
solutions $ \gamma_\omega$ should then minimize the free energy
\begin{equation}
F=-T\sum_{\omega>0} ln [A^{2}(\omega)+
4\gamma_{\omega}^{2}\omega^{2}(1+a(\omega))^{2}]+
\frac{\gamma_{\omega}^2}{2U}
+\tilde{\epsilon}.
\end{equation}
Here we use the notations:
\begin{equation}
A(\omega)=\tilde{\epsilon}^2+\omega^{2}(1+a(\omega))^{2}
-\gamma_{\omega}^{2}-b^{2}(\omega)|\Delta|^{2}\cos^{2}(\frac{\phi}{2}),
\end{equation}
\begin{equation}
a(\omega)= \Gamma\frac{\sqrt{\omega^2 +|\Delta|^{2}}}
{\omega^2},
\end{equation}
where $\Gamma=2 \pi t^2 N(0)$ is the bare impurity level width,
 $b(\omega)=\Gamma / \omega$ , and $\phi$ is the phase difference
between the two superconductors.
The self-consistent equation for $\gamma_{\omega}$ is
 
\begin{equation}
\frac{1}{2U}-2T\sum_{\omega>0} 
\frac{2 \omega^{2} (1+a(\omega))^{2}
-A(\omega)}{A^{2}(\omega)+
4\gamma_{\omega}^{2}\omega^{2}(1+a(\omega))^{2}} =0. 
\end{equation}
Once the solutions are defined we can calculate the
current $J=(2e/\hbar)\partial F/\partial\phi$ and the impurity
occupancy $n=\partial F/\partial\epsilon_{0}$,

\begin{equation}
J
=-(2e/\hbar)\sin(\phi)T\sum_{\omega>0}\frac{|\Delta|^{2}
b^{2}(\omega)A(\omega)}{A^{2}(\omega)+
4\gamma_{\omega}^{2}\omega^{2}(1+a(\omega))^{2}},
\end{equation}
\begin{equation}
n
=1-4T\sum_{\omega>0}\frac{\tilde{\epsilon}
A(\omega)}{A^{2}(\omega)+
4\gamma_{\omega}^{2}\omega^{2}(1+a(\omega))^{2}}. \nonumber
\end{equation}

We now analyze the main results of the present study.
All the parameters having the dimension of energy
($\epsilon,U,\Gamma,T$) are expressed in units of $|\Delta|$ 
and the current is given in units of $|\Delta| e /\hbar$.
\begin{figure} [tbp]
\centerline{\epsfxsize=7.0cm \epsfbox{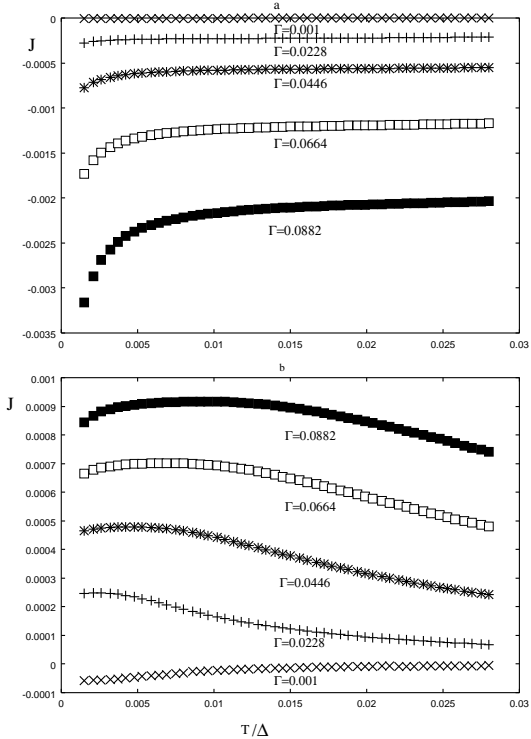}}
\caption{Dependence of the Josephson current on 
temperature 
for $U=2.1$ (a) and $U=3.0$ (b) for different transparencies $\Gamma$. 
Note the sign reversal for strong $U$.
Here $\epsilon_{0}=-2.0$ and $\phi=\pi/12$ are fixed. The units of all 
quantities are explained in the text.}
\end{figure}
\noindent
In Fig.1a,b, 
the current is displayed versus temperature in
the low temperature region and for coupling strengths 
$\Gamma$ ranging between $0.001$ and $0.0882$. 
At small Hubbard interaction 
($U$=2.1, Fig.1a) the
junction is in a '$\pi$-state for which the current 
(within the present geometry of the
tunneling direction) is negative. The current is strongly
dependent on temperature, which is markedly distinct from the classical
Ambegaokar-Baratoff formula. It is typical
for superconducting systems for which the zero energy mid-gap bound state 
plays the major role. At higher values of $U$ (Fig. 1b)
the pattern is inverted:
The current for most coupling strengths is now
 positive and the '$\pi$' junction is transformed into a '$0$'
junction  with slightly weaker temperature dependence. 
A similar situation takes place also in $s$-wave superconductors
\cite{spi,aro}, where it is attributed 
to the single occupancy of the impurity which 
then becomes a degenerate magnetic moment. We 
have thus presented another
 example for this scenario, though the temperature
dependence is quite different.
\begin{figure} [tbp]
\centerline{\epsfxsize=7.0cm \epsfbox{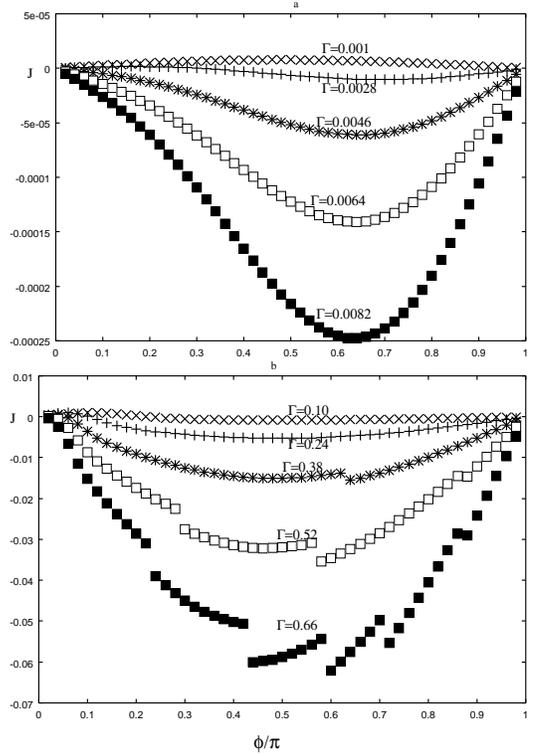}}
\caption{Dependence of the Josephson current on the phase difference 
for (a) small transparencies and intermediate temperature ($T=0.01$) and 
(b) large transparencies and very low 
temperature ($T=0.001$). Other parameters 
are $\epsilon_{0}=-2.0$ and $U=2.6$.}
\end{figure}

\noindent
Now let us discuss the behavior of the current $J(\phi)$ as
function of the phase difference as displayed in Figs.2a,b 
for fixed $U$ and $T$ and for numerous transparencies $\Gamma$.
Once again it is clear that as the transparency is increased,
the sign of the current is reversed. Moreover, at lower temperatures 
(Fig.2b) when $\Gamma/\Delta \geq 0.4$ it becomes
irregular with jumps at certain values of $\phi$ at which the impurity  
is nearly incompressible. This effect can manifest itself 
when the flowing current $I>I_{c}$. In the frame of the
RSJ model the averaged voltage $\bar{V}$ on the junction is 
related to the current $I$ and the resistance $R$ as, 

\begin{equation}
R=\bar{V} \int_{0}^{2\pi}\frac{d\phi}{2\pi (I-J(\phi))}.
\end {equation}
The irregular pattern of $J(\phi)$ results in a deviation of
the $\bar{V}(I)$ characteristic from its classical expression
$I=\sqrt{J^{2}_{c}+(\bar{V}/R)^2 }$. Such deviation is  seen on
Fig.3 where the
 difference is mainly exhibited at low voltage.
\begin{figure} [tbp]
\centerline{\epsfxsize=7.0cm \epsfbox{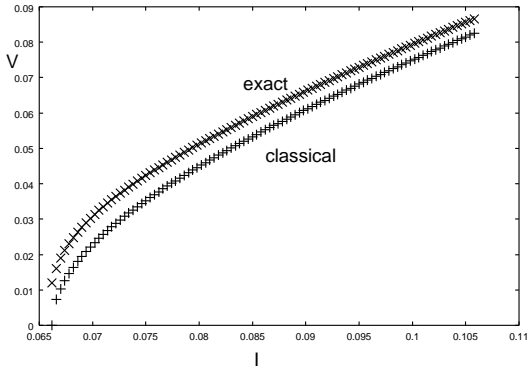}}
\caption{$V(I)$ characteristics of the $p$ wave junction (upper curve) 
compared with the classical square root expression (lower curve). The relevant 
parameters are $\epsilon_{0}=-2.0,U=3.0$ and $\Gamma=0.66$.}
\end{figure}

To summarize, we have solved the problem of transport between 
two unconventional superconductors through an impurity 
and traced the dependence of the current on
temperature, Coulomb interaction barrier transparency 
and phase of condensates. 
The essentially non-perturbative,
self-consistent approach we have used yields a finite value for the
current without adding relaxation terms
that should be included if perturbation approach is adopted.
It is 
mandatory at this final stage to point out the peculiarities 
related to the physics of non $s$-wave superconductors. 

\noindent
1. The contribution to the current in our case originates principally from
the surface bound state which is related to the asymmetry of the pair 
potential, and has no analog for $s$-wave superconductors. The
contribution of this state in low transparency
junctions results in a large current, that is,
$|J_{p}/J_{s}|\sim \sqrt{\Delta/\Gamma}$. It is also marked by a stronger
temperature dependence especially at low temperatures.

\noindent
2. For unconventional superconductors, the Josephson tunneling
through an interacting
quantum dot can serve as an indicator to distinguish odd parity
superconductors from even parity ones. 
Superconductors with both $p$ 
and $d$-wave symmetry of the order parameter
have surface bound state which 
contributes to the current (mainly at low temperatures).
Yet, as we indicated above, for even-symmetry 
($d$-wave) superconductors, the current vanishes
in the limit of point-like impurity.
In sharp distinction, under the same conditions, the current 
is maximal for odd-symmetry ($p$-wave) superconductors.

\noindent
{\bf Acknowledgment}: This research is supported by grants from the 
Israeli Science Foundation ({\em Center of Excellence and 
Non-Linear Tunneling}), The German-Israeli 
DIP foundation and the US-Israel BSF foundation.   
 

\end{multicols}
\end{document}